\documentclass{article}%
\usepackage{amsmath}
\usepackage{amsfonts}
\usepackage{amssymb}
\usepackage{graphicx}%
\providecommand{\U}[1]{\protect\rule{.1in}{.1in}}

\begin{document}

\begin{center}
\bigskip{\Huge Overlapping resonance branches of a gauge-invariant
Lorentz-violating massive vector}

{\Huge \bigskip}

{\Huge \bigskip}

\textbf{Z. Kepuladze}\footnote{zurab.kepuladze.1@iliauni.edu.ge,
zkepuladze@yahoo.com}

\bigskip

\textit{Institute of Theoretical Physics, Ilia State University, 0162 Tbilisi,
Georgia\ \vspace{0pt}\\[0pt]}

\textit{and} \textit{Andronikashvili} \textit{Institute of Physics, 0177
Tbilisi, Georgia\ }

\bigskip

\bigskip

\textbf{Abstract}
\end{center}

We investigate the propagation and resonance behavior of an Abelian massive
vector field in the presence of a gauge-invariant, CPT-even, dimension-four
Lorentz-violating kinetic operator constructed from a single preferred
four-vector. In the massless theory, the propagator contains an additional
algebraic pole that decouples from conserved currents, leaving two physical
degrees of freedom. After spontaneous breaking of the internal U(1) symmetry,
all three vector polarizations become physical but separate into two
orthogonal dispersion branches carrying two and one physical polarization states,
respectively. We show that this decomposition is preserved under Dyson
resummation for a transverse matter vacuum polarization and calculate the
corresponding decay rates and fermion-annihilation amplitude. The split-pole
structure and associated double resonance behavior cannot be understood through a finite number of perturbative Lorentz-violating insertions. The contribution of the second branch in physical processes is strongly dependent on the
initial momentum geometry: it vanishes for massless head-on fermions in a
timelike background, whereas boosted nearly parallel configurations can
generate an energy-enhanced correction. Spacelike and lightlike backgrounds
additionally lead to directional and potentially sidereal variations of the
resonance signal.

\thispagestyle{empty}\newpage

\section{Introduction}

A broad range of possible violations of Lorentz invariance in quantum
electrodynamics \cite{Nielsen-Chadha,chk,chk1,kost0,kost,chk2,chk3} and
related gauge theories has been investigated
\cite{tomb,kost1,kost2,rizzo,chk4,kost3,chk5,chk6,chk7,chk8,chk9}. General
parametrizations of Lorentz-violating operators have been developed, their
one-loop renormalization properties have been studied, and numerous physical
effects have been derived and constrained experimentally
\cite{CFJ,SME-K1,SME-K2,Coleman-Glashow,Gagnon-Moore,kost4,Kostelecky-Russell,LHAASO-LIV,Satunin,Renorm1,Renorm2,Renorm3}%
. The propagating degrees of freedom, dispersion relations, stability, and
consistency conditions of different Lorentz-violating vector theories have
also been analyzed \cite{St,St1,St2,St3}.

In most treatments, Lorentz invariance violation (LIV) in the vector sector is introduced
through explicitly gauge-invariant operators. This is a well-motivated
assumption, since gauge invariance strongly constrains the allowed
interactions and radiative corrections. Nevertheless, it has been demonstrated
that the preservation of gauge invariance may have nontrivial limitations. In
particular, the one-loop photon self-energy does not always remain gauge invariant
\cite{Renorm4,Renorm5,Renorm6}.

The complete conditions under which gauge invariance is maintained at higher
orders in the Lorentz-violating parameters and beyond the leading loop
approximation are therefore model dependent and are not automatically
guaranteed by the manifestly gauge-invariant form of the classical action.
Moreover, the Ward--Takahashi identities for QED involving the fermion
self-energy and the fermion--vector vertex primarily constrain the
conservation of the fermionic vector current.

The relation between physical Lorentz violation and gauge invariance is
consequently subtle. A Lorentz-violating quadratic operator that is not gauge
invariant may, under certain conditions, amount only to a noncovariant
gauge-fixing term rather than to a genuine modification of the physical vector
dynamics. Additional theoretical input is then required to distinguish
physical Lorentz violation from a particular gauge fixing. Even when a
phenomenological LIV operator is explicitly gauge invariant at the classical
level, the conditions under which gauge invariance is preserved by quantum
corrections are model dependent and not completely established
\cite{Renorm1,Renorm2,Renorm3,Renorm4,Renorm5,Renorm6}. We therefore restrict
the present analysis to the perturbative regime in which the
vacuum-polarization tensor retains its standard transverse form.

The behavior of massless and massive vector modes in the presence of Lorentz
violation is also qualitatively different. The massless vector field retains
two physical propagating degrees of freedom, although their dispersion
relation may be modified. In the massive theory, the different polarization
modes may instead propagate according to distinct dispersion relations,
leading to a splitting of the otherwise degenerate massive-vector modes.
Below, we assume that the preferred LIV direction in spacetime is fixed by the vector $n_{\mu}$. For timelike and spacelike $n_{\mu}$, $\left\vert
n_{\mu}n^{\mu}\right\vert =1$, while for a lightlike vector $\left\vert n_{\mu}n^{\mu
}\right\vert =0$, with normalization $\left\vert n_{0}\right\vert =\left\vert
\overrightarrow{n}\right\vert =1$. We introduce a gauge invariant Lorentz-violating
contribution to the kinetic operator of an Abelian U(1) vector field and
reconsider several known features, as well as some additional consequences, of
the vector-field dynamics in the U(1)-symmetric and spontaneously broken phases.

\section{Gauge-invariant LIV dynamics of massless and massive vectors}

With the vector $n_{\mu}$ fixing the LIV background, the gauge-invariant,
CPT-even, dimension-four quadratic contribution constructed from a single
preferred vector that modifies the vector-field dispersion relation is given
by:%
\begin{equation}
\Delta L_{LIV}=\frac{\delta}{4}n^{\mu}F_{\mu\nu}n_{\lambda}F^{\lambda\nu}
\label{LIVop}%
\end{equation}
where $\delta$ is the LIV parameter and $F_{\mu\nu}$ is the field strength
tensor of the vector field $A_{\mu}$. Even at a surface level, it is
immediately clear that (\ref{LIVop}) modifies the dispersion relation of the
massless vector:%
\begin{equation}
k^{2}-\delta k_{n}^{2}=0
\end{equation}
with $k_{\mu}$ denoting the four-momentum of the field, $k^{2}=k_{\mu}k^{\mu}%
$, and $k_{n}=n_{\mu}k^{\mu}$. Applying the Landau gauge condition
$\partial_{\mu}A^{\mu}=0$, the propagator takes the form
\begin{equation}
D_{\mu\nu}=\frac{-i}{k^{2}-\delta k_{n}^{2}}\left[  g_{\mu\nu}-\frac{k_{\mu
}k_{\nu}}{k^{2}}+\frac{\delta\overline{n}_{\mu}\overline{n}_{\nu}}{1-\delta
n^{2}}\right]  \text{, \ \ \ \ \ \ \ \ }\overline{n}_{\mu}=n_{\mu}-k_{\mu
}k_{n}/k^{2}%
\end{equation}
which is manifestly orthogonal to the momentum, and at the same time
orthogonal to $n_{\mu}$ on the mass shell. A similar result follows in the
axial gauge $n_{\mu}A^{\mu}=0$, where the propagator is manifestly orthogonal
to $n_{\mu}$ and becomes orthogonal to the momentum only on the mass shell.
The Lorentz invariant limit is recovered properly.

An interesting detail emerges if we define the projectors
\begin{equation}
P_{k}^{\mu\nu}=\frac{k^{\mu}k^{\nu}}{k^{2}}\text{, \ \ \ \ \ }P_{n}^{\mu\nu
}=\frac{\overline{n}^{\mu}\overline{n}^{\nu}}{\overline{n}^{2}}%
\end{equation}
the propagator can then be rewritten as%
\begin{equation}
iD^{\mu\nu}=\frac{g^{\mu\nu}-P_{k}^{\mu\nu}-P_{n}^{\mu\nu}}{k^{2}-\delta
k_{n}^{2}}+\frac{P_{n}^{\mu\nu}}{(1-\delta n^{2})k^{2}}%
\end{equation}
Taking the trace yields
\begin{equation}
ig_{\mu\nu}D^{\mu\nu}=\frac{2}{k^{2}-\delta k_{n}^{2}}+\frac{1}{(1-\delta
n^{2})k^{2}}%
\end{equation}

This indicates, in principle, the presence of two poles: one at $k^{2}-\delta
k_{n}^{2}=0$ and another at $k^{2}=0$. One might conclude that the first pole
corresponds to two propagating degrees of freedom and the second to one.
However, the pole at $k^{2}=0$ is not physical, as it disappears when
sandwiched between conserved fermion currents. Thus, in truth, only two
degrees of freedom propagate.

If, for this Abelian vector field $A_{\mu}$, the standard Higgs mechanism is
arranged and spontaneous breaking of the internal symmetry by a complex scalar
field takes place, then without introducing any additional LIV source the
vector field absorbs the scalar Goldstone mode and acquires a mass $M$. Since
gauge invariance is formally maintained, the unitary gauge can be chosen as
the simplest alternative. In this case, the propagator in the unitary gauge
takes a form similar to the massless case, but with an altered pole structure:%
\begin{equation}
iD^{\mu\nu}=\frac{g^{\mu\nu}-P_{k}^{\mu\nu}-P_{n}^{\mu\nu}}{k^{2}-\delta
k_{n}^{2}-M^{2}}+\frac{P_{n}^{\mu\nu}}{(1-\delta n^{2})k^{2}-M^{2}}%
-\frac{P_{k}^{\mu\nu}}{M^{2}}%
\end{equation}
Now both poles are physical, indicating that different degrees of freedom
(different polarizations) propagate differently. On the LIV shell
$k^{2}-\delta k_{n}^{2}-M^{2}=0$, two degrees of freedom propagate, while on
the second shell, which is Lorentz invariant in form, $(1-\delta n^{2}%
)k^{2}-M^{2}=0$, one degree of freedom propagates. Importantly, these
different degrees of freedom do not mix and remain orthogonal. In effect, the LIV massive-vector multiplet exhibits a splitting of its dispersion relations, which also corresponds to a mass splitting except in the purely timelike case, with the orthogonal components propagating independently.

The propagator of such an LIV massive vector is thus split into two distinct
and orthogonal sectors. However, the same current interacts with both, so in
any physical process they contribute as two separate mediators summed
together. Consequently, such processes should exhibit two distinct resonances
corresponding to the two poles. Since this is an LIV effect, the resonance
locations lie very close to each other, producing an overlapping signal.
Naturally, in the Lorentz invariant limit $\delta\rightarrow0$, the standard
propagator is restored, the dispersion relation splitting disappears, and pole
degeneracy between different vector modes is recovered.

Because the two sectors remain orthogonal and propagate according to different
dispersion relations, their decay rates should be calculated independently
and, in general, need not coincide. Next, we will calculate how different they
are and show that the orthogonal decomposition is preserved under Dyson
resummation when the matter vacuum polarization has the usual transverse form.

\subsection{Loop corrections and the two resonance branches}

For Lorentz-invariant fermions with vector couplings, as well as for the
gauge-invariant sum of the scalar bubble and seagull contributions, the
vacuum-polarization tensor has the usual transverse form,
\begin{equation}
\Pi^{\mu\nu}(k)=\left(  g^{\mu\nu}-k^{\mu}k^{\nu}/k^{2}\right)  \Pi(k^{2})
\end{equation}
then we can check that
\begin{equation}
D^{\mu\nu}\Pi_{\nu}^{\lambda}(k)=\frac{g^{\mu\lambda}-P_{k}^{\mu\lambda}%
-P_{n}^{\mu\lambda}}{k^{2}-\delta k_{n}^{2}-M^{2}}\Pi(k^{2})+\frac{P_{n}%
^{\mu\lambda}}{(1-\delta n^{2})k^{2}-M^{2}}\Pi(k^{2})
\end{equation}
and Dyson resummation gives
\begin{equation}
\overline{D}^{\mu\lambda}=\frac{g^{\mu\lambda}-P_{k}^{\mu\lambda}-P_{n}%
^{\mu\lambda}}{k^{2}-\delta k_{n}^{2}-M^{2}-\Pi(k^{2})}+\frac{P_{n}%
^{\mu\lambda}}{(1-\delta n^{2})k^{2}-M^{2}-\Pi(k^{2})}-\frac{P_{k}^{\mu
\lambda}}{M^{2}}%
\end{equation}
As usual $\Pi(k^{2})$ renormalizes the mass parameter. Near the resonance poles, it also acquires an imaginary part determined by the corresponding decay rate. Although the same transverse self-energy function appears in both denominators, the two widths need not be equal. The self-energy is evaluated on different dispersion branches, and the corresponding pole-normalization factors are different. Through the optical theorem, the imaginary part of each pole function is related to the decay rate of the corresponding polarization state. Adopting the form motivated in the \cite{Z-res}, we can write
\begin{align}
\overline{D}^{\mu\lambda}  &  =\frac{g^{\mu\lambda}-P_{k}^{\mu\lambda}%
-P_{n}^{\mu\lambda}}{k^{2}-(M_{1,eff}-ik_{0}\Gamma_{1eff}(k)/2M_{1,eff})^{2}%
}+\nonumber\\
&  \frac{P_{n}^{\mu\lambda}}{k^{2}-(M_{2,eff}-ik_{0}\Gamma_{2,eff}%
(k)/2M_{2,eff})^{2}}-\frac{P_{k}^{\mu\lambda}}{M^{2}} \label{pro}%
\end{align}
where we detone by $\Gamma_{1,2eff}(k)$ corresponding decay rates of the
orthogonal degrees of freedom propagating with momentum $k_{\alpha}$,
\begin{align}
M_{1,eff}^{2}  &  =M^{2}+\delta k_{n}^{2}\label{LV}\\
M_{2,eff}^{2}  &  =M^{2}+\delta n^{2}k^{2} \label{LI}%
\end{align}

Finally, the split-pole structure is not visible if the LIV quadratic operator
of the style of (\ref{LIVop}) is treated only as a finite number of
perturbative insertions. At any finite order, an expansion of the form
\begin{equation}
\frac{1}{P_{0}(k)-\delta X(k)}=\frac{1}{P_{0}(k)}+\delta\frac{X(k)}{P_{0}%
^{2}(k)}+\mathcal{O}(\delta^{2})
\end{equation}
retains the unperturbed pole location. To resolve the separate LIV branches
and the associated two-resonance structure, the quadratic LIV operator must be
incorporated into the exact propagator or resummed to all orders.

\subsection{Decay rate}

For brevity, let us assume that the massive vector decays only via a
fermion-antifermion channel (i.e. the scalar decay channel is not
kinematically viable), and that fermion masses are negligible compared to the
boson mass. Denoting $e$ as the fermion charge, the squared matrix element for
the boson decay is
\begin{equation}
\left\vert \mathcal{M}\right\vert ^{2}=e^{2}\xi_{\mu}\xi_{\nu}Tr[\not p
\gamma^{\mu}\not q  \gamma^{\nu}]=4e^{2}\xi_{\mu}\xi_{\nu}(2p^{\mu}q^{\nu
}-g^{\mu\nu}(p_{\lambda}q^{\lambda}))
\end{equation}
where $\xi_{\mu}$ is the polarization vector, $p^{\mu}$ and $q^{\mu}$ are the
final fermion momenta, and $k^{\mu}=p^{\mu}+q^{\mu}$ is the boson momentum. In
the massless-fermion limit, energy-momentum conservation and integration over
the final-state phase space allow the replacements
\begin{equation}
p_{\lambda}q^{\lambda}=\frac{k^{2}}{2}=\frac{M_{eff}^{2}}{2},\text{
\ \ }p^{\mu}q^{\nu}\rightarrow\frac{1}{12}\left(  M_{eff}^{2}g^{\mu\nu
}+2k^{\mu}k^{\nu}\right)  \label{Replace}%
\end{equation}

Thus the matrix element becomes
\begin{equation}
\left\vert \mathcal{M}\right\vert ^{2}=-\frac{4e^{2}M_{eff}^{2}}{3}\xi_{\mu
}\xi^{\mu}%
\end{equation}
where the mass-shell condition $k_{\mu}\xi^{\mu}=0$, valid for all physical
polarizations, has been used. \ Also using that $\xi_{\mu}\xi^{\mu}=-1$, we
obtain for each physical degree of freedom with its characteristic dispersion
relation%
\begin{equation}
\left\vert \mathcal{M}_{i}\right\vert ^{2}=\frac{4e^{2}M_{i,eff}^{2}}%
{3}\text{, \ }i=1,2
\end{equation}
Here\ $i=1$ labels the two degenerate polarizations on the first branch with
dispersion relation (\ref{LV}), whereas \ $i=2$ labels the single-polarization
branch with dispersion relation (\ref{LI}).

The corresponding decay width is
\begin{equation}
\Gamma_{i,eff}=\frac{1}{2k_{0}}\int\frac{d^{3}p}{(2\pi)^{3}2p_{0}}\frac
{d^{3}q}{(2\pi)^{3}2q_{0}}(2\pi)^{4}\delta^{(4)}(k-p-q)|\mathcal{M}_{i}%
|^{2}\nonumber
\end{equation}
\footnote{Strictly speaking, for each Lorentz-violating dispersion branch, the
conventional normalization factor $\dfrac{1}{2k_{0}}$ must be replaced by the
corresponding on-shell Jacobian, namely $2k_{0}\rightarrow\left\vert
\partial_{k_{0}}\left(  Dispersion-Relation\right)  \right\vert $. The
resulting corrections are of order $\delta$, are not enhanced at high
energies, and are numerically insignificant for the effects considered here;
they will therefore be neglected.}which evaluates to%
\begin{equation}
\Gamma_{i,eff}=\frac{|\mathcal{M}_{i}|^{2}}{16\pi k_{0}}=\frac{e^{2}%
M_{i,eff}^{2}}{12\pi k_{0}} \label{DecayR}%
\end{equation}

If there are $N$ fermion species into which the vector boson can decay, then a
factor of $N$ should appear in the total decay rate.

\bigskip

\subsection{Resonance behavior}

Taking into account (\ref{DecayR}) in the propagator (\ref{pro}), we obtain%
\begin{align}
\overline{D}^{\mu\lambda}(k)  &  =\frac{g^{\mu\lambda}-P_{k}^{\mu\lambda
}-P_{n}^{\mu\lambda}}{k^{2}-M_{1,eff}^{2}(1-i\overline{e}^{2})^{2}%
}+\nonumber\\
&  \frac{P_{n}^{\mu\lambda}}{k^{2}-M_{2,eff}^{2}(1-i\overline{e}^{2})^{2}%
}-\frac{P_{k}^{\mu\lambda}}{M^{2}}%
\end{align}
where $\overline{e}^{2}\equiv Ne^{2}/24\pi$. When considering annihilation (or
more generally scattering) between two fermions, this propagator must be
sandwiched between the corresponding currents. Denoting $k^{\mu}$ as the
four-momentum carried by the vector mediator (i.e. the total energy-momentum
of the system), the matrix element reads
\begin{equation}
\mathcal{M}_{el}=J_{1\mu}(k)\overline{D}^{\mu\lambda}(k)J_{2\lambda}(k)
\end{equation}
Using $k^{\mu}J_{1,2\mu}(k)=0$, this reduces to%
\begin{equation}
\mathcal{M}_{el}=\frac{J_{1\mu}(k)\left(  g^{\mu\lambda}-P_{n}^{\mu\lambda
}\right)  J_{2\lambda}(k)}{k^{2}-M_{1,eff}^{2}(1-i\overline{e}^{2})^{2}}%
+\frac{J_{1\mu}(k)P_{n}^{\mu\lambda}J_{2\lambda}(k)}{k^{2}-M_{2,eff}%
^{2}(1-i\overline{e}^{2})^{2}}\nonumber
\end{equation}

\bigskip After algebraic rearrangement:%

\begin{equation}
\mathcal{M}_{el}=\frac{J_{1}^{\nu}(k)J_{2}^{\mu}(k)}{k^{2}-M_{1,eff}%
^{2}(1-i\overline{e}^{2})^{2}}\left[  g_{\mu\nu}+\frac{n_{\mu}n_{\nu
}(1-i\overline{e}^{2})^{2}\delta k^{2}}{(k^{2}-M_{2,eff}^{2}(1-i\overline
{e}^{2})^{2})}\right]
\end{equation}
In the $J_{1n}(k)J_{2n}(k)$ part, zeroth-order effects between the two poles
cancel, leaving only LIV contributions linear in $\delta$, as expected, since
pole splitting does not occur in the LI limit.

Defining the usual resonance factors,%

\begin{align}
R_{s1}  &  =\frac{k^{4}}{(k^{2}-M_{1,eff}^{2}(1-\overline{e}^{4}%
))^{2}+4\overline{e}^{4}M_{1,eff}^{4}}\label{R1}\\
R_{s2}  &  =\frac{1}{(k^{2}-M_{2,eff}^{2}(1-\overline{e}^{4}))^{2}%
+4\overline{e}^{4}M_{2,eff}^{4}} \label{R2}%
\end{align}
and denoting $p_{1,2}^{\mu}$ as the initial fermion momenta, we find%

\begin{align}
\left\vert \mathcal{M}_{el}\right\vert ^{2}  &  =\frac{16e^{4}}{3}%
R_{s1}[1+\delta R_{s2}(n^{2}k^{2}-4p_{1n}p_{2n})F(k^{2},k_{n})]\\
&  F(k^{2},k_{n})=(1-\overline{e}^{4})\left(  k^{2}-M_{2,eff}^{2}%
(1-\overline{e}^{4})\right)  -\nonumber\\
&  4\overline{e}^{4}M_{2,eff}^{2}+\frac{\delta(1+\overline{e}^{4})^{2}\left(
n^{2}k^{2}-k_{n}^{2}\right)  }{2}%
\end{align}
Here, momentum conservation and replacement (\ref{Replace}) have been used.
The resonance factor $R_{s1}$ amplifies the entire expression, while $R_{s2}$
enhances the second and third terms. The second term represents interference
with the LI amplitude, whereas the explicitly quadratic $\delta^{2}$ term
corresponds to the squared modulus of the pole splitting contribution. This
algebraic interference should not be confused with interference between
orthogonal polarization sectors.

Although the terms carry explicit $\delta$ and $\delta^{2}$ factors, $R_{s2}$
contains the exact second denominator, so this is not a finit order
expansion near resonance. Naturally, we assume that the coupling constant $e$
is a small perturbative parameter; otherwise, the analysis would not be meaningful.

The second resonance occurs at
\begin{equation}
k_{res2}^{2}=M_{2,eff}^{2}(1-\overline{e}^{4})\text{ \ }\rightarrow\text{
}k_{res2}^{2}=\frac{M^{2}(1-\overline{e}^{4})\text{ }}{\text{\ }(1-\delta
n^{2}(1-\overline{e}^{4}))}\approx M^{2}(1-\overline{e}^{4}) \label{Re2}%
\end{equation}
which is nearly the LI result.

At this resonance, the squared amplitude becomes
\begin{align}
\left\vert \mathcal{M}_{el}\right\vert _{res2}^{2}  &  \approx\frac{16e^{4}%
}{3}R_{s1}[1-\delta\left(  n^{2}-4p_{1n}p_{2n}/M^{2}\right)  +\nonumber\\
&  \frac{\delta^{2}}{8\overline{e}^{4}}\left(  n^{2}-\left(  k_{n}^{2}\right)
_{res2}/M^{2}\right)  \left(  n^{2}-4p_{1n}p_{2n}/M^{2}\right)  ] \label{A1}%
\end{align}

The impact of the second resonance depends strongly on $\delta$, on the ratio
$\delta^{2}/(8e^{4})$, and on the initial momentum configuration. For example,
if $n_{\mu}$ is timelike and the initial fermion energies are $E_{1}$ and
$E_{2}$ ($E=E_{1}+E_{2}$), then%

\begin{align}
\left\vert \mathcal{M}_{el}\right\vert _{res2}^{2}  &  \approx\frac{16e^{4}%
}{3}R_{s1}[1-\delta\left(  1-4E_{1}(E_{res2}-E_{1})/M^{2}\right)  +\nonumber\\
&  \frac{\delta^{2}}{8\overline{e}^{4}}\left(  1-E_{res2}^{2}/M^{2}\right)
\left(  1-4E_{1}(E_{res2}-E_{1})/M^{2}\right)  ] \label{A2}%
\end{align}
At high energies compared to the boson mass, the leading contribution is
\begin{equation}
\left\vert \mathcal{M}_{el}\right\vert _{res2}^{2}\approx\frac{16e^{4}}%
{3}R_{s1}[1+4\delta\frac{E_{1}(E_{res2}-E_{1})}{M^{2}}+\frac{\delta^{2}%
}{2\overline{e}^{4}}\frac{E_{res2}^{2}}{M^{2}}\frac{E_{1}(E_{res2}-E_{1}%
)}{M^{2}}]
\end{equation}
which maximizes for $E_{1}=E_{2}=E_{res2}/2$:%
\begin{equation}
\left\vert \mathcal{M}_{el}\right\vert _{res2}^{2}\approx\frac{16e^{4}}%
{3}R_{s1}[1+\delta\frac{E_{res2}^{2}}{M^{2}}+\frac{\delta^{2}}{8\overline
{e}^{4}}\frac{E_{res2}^{4}}{M^{4}}]
\end{equation}

The linear $\delta$ term contributes at the same order as the LIV modification
in $R_{s1}$, while the $\delta^{2}$ term is parametrically suppressed by the
small LIV parameter but can be enhanced by the second pole resonance and by
strong energy amplification.

To illustrate the possible magnitude of the effect at an electroweak mass
scale, consider a $Z$-boson mass benchmark with $M\approx90\ \text{GeV}$,
laboratory-frame resonance energy $E_{res2}=10\ \text{TeV (an LHC-scale
energy)}$, and a representative value $\overline{e}^{2}\sim1/137$. Then
\begin{equation}
\left\vert \mathcal{M}_{el}\right\vert _{res2}^{2}\approx\frac{16e^{4}}%
{3}R_{s1}[1+\delta\cdot10^{4}+\left(  \delta\cdot10^{6}\right)  ^{2}]
\end{equation}
Thus, for $\delta\sim10^{-8}$, the linear and quadratic corrections in
$\delta$ are comparable and are both of order $10^{-4}$. Previous estimates
indicate that LIV parameters of this order may produce detectable imprints in
high-energy collider measurements \cite{Z-res,Z-res1}. At higher
energies, the quadratic contribution grows rapidly, scaling as the fourth
power of the laboratory-frame resonance energy.

These optimistic estimates require a specific momentum configuration. The
incoming momenta $p_{1,2}$ must not be antiparallel but nearly parallel, with
an opening angle approximately given by%

\begin{equation}
\theta\approx\frac{2M}{E_{res2}}%
\end{equation}

Although the energy chosen in this illustrative example is of the order
accessible at modern colliders, the nearly parallel momentum configuration
that maximizes the effect is not the standard leading-order partonic geometry
of a hadron collider and should be regarded here only as a kinematic benchmark.

The standard collider geometry is instead head-on, with antiparallel incoming
parton momenta. In this case, the resonance condition reduces to%

\begin{equation}
M^{2}\approx4E_{1}(E_{res2}-E_{1}) \label{con1}%
\end{equation}
which cancels all the contributions from the second resonance pole in the
leading order and simplifies (\ref{A2}) to
\begin{equation}
\left\vert \mathcal{M}_{el}\right\vert _{res2}^{2}\approx\frac{16e^{4}}%
{3}R_{s1}%
\end{equation}
\qquad{}

In this configuration, the explicit contribution associated with the second pole vanishes, at leading order. Indeed, $4n^{2}p_{1n}p_{2n}/M^{2}\approx1$, which is
(\ref{con1}). The remaining dependence on LIV is contained exclusively in
$R_{s1}$.

For spacelike or lightlike $n_{\mu}$, the result becomes direction dependent,
and this exact cancellation is not generic. In an Earth-based experiment, the
orientation of the collision axis relative to a fixed preferred direction
generally changes with sidereal time, producing a corresponding modulation of
the signal similar to that discussed in \cite{Z-res1}.\ Let $\alpha_{L}(t)$
denote the angle between a spacelike preferred direction and the collision
axis. For massless head-on incoming fermions,
\begin{align*}
\left(  n^{2}-4p_{1n}p_{2n}/k^{2}\right)  _{res2}  &  \approx-\sin^{2}%
\alpha_{L}\\
\left(  n^{2}-\left(  k_{n}^{2}\right)  _{res2}/k^{2}\right)  _{res2}  &
\approx-\sin^{2}\alpha_{L}-\frac{E_{res2}^{2}}{M^{2}}\cos^{2}\alpha_{L}%
\approx-\frac{E_{res2}^{2}}{M^{2}}\cos^{2}\alpha_{L}%
\end{align*}
assuming $\sin^{2}\alpha_{L}\ll\frac{E_{res2}^{2}}{M^{2}}\cos^{2}\alpha_{L}$.
At the second resonance defined by (\ref{Re2}), eq. (\ref{A1}) gives%
\begin{equation}
\left\vert \mathcal{M}_{el}\right\vert _{res2}^{2}\approx\frac{16e^{4}}%
{3}R_{s1}[1+\delta\sin^{2}\alpha_{L}+\frac{\delta^{2}}{8\overline{e}^{4}}%
\frac{E_{res2}^{2}}{M^{2}}\sin^{2}\alpha_{L}\cos^{2}\alpha_{L}]
\end{equation}
Thus, only the quadratic contribution receives a high-energy enhancement, and
even this enhancement scales as $\dfrac{E_{res2}^{2}}{M^{2}}$, rather than as
$\dfrac{E_{res2}^{4}}{M^{4}}$. This standard collision configuration is
therefore not favorable for obtaining a large rate enhancement from the second
resonance. Nevertheless, its directional and sidereal dependence may provide a
distinctive LIV signature. For a lightlike preferred vector, the exact timelike cancellation is likewise absent in general. The energy scaling is similar to that in the spacelike case, but the detailed angular dependence differs.

Finally, the final-state-integrated annihilation cross section for fixed
initial momenta can be expressed as
\begin{equation}
\sigma_{LIV}=\sigma_{0LI}R_{s1}[1+\delta R_{s2}(n^{2}k^{2}-4p_{1n}%
p_{2n})F(k^{2},k_{n})]
\end{equation}
where $\sigma_{0LI}$ denotes the unpolarized tree-level cross section for the
same fermionic process, with identical couplings and external-state
kinematics, but mediated by a massless Lorentz-invariant vector field.
Consequently,%
\[
\lim_{\delta\rightarrow0}\sigma_{LIV}=\lim_{\delta\rightarrow0}\left(
\sigma_{0LI}R_{s1}\right)
\]
reproduces the usual Lorentz-invariant massive-vector Breit--Wigner cross section.

\section{Conclusion}

We have studied the propagation and resonance behavior of an Abelian vector
field in the presence of a gauge-invariant, CPT-even, dimension-four
Lorentz-violating kinetic operator constructed from a single preferred
four-vector. In the unbroken phase, the propagator admits an algebraic
decomposition containing two pole structures. However, the additional pole
decouples when the propagator is contracted with conserved currents, and only
the usual two physical massless-vector degrees of freedom propagate.

After spontaneous breaking of the internal $U(1)$ symmetry, the situation
changes qualitatively. The three physical polarizations of the massive vector
separate into two orthogonal sectors: two polarizations propagate on the
dispersion branch
\[
k^{2}-\delta k_{n}^{2}-M^{2}=0,
\]
whereas the remaining polarization propagates on
\[
(1-\delta n^{2})k^{2}-M^{2}=0.
\]
Thus, the Lorentz-violating massive vector gives rise to two nearby physical
resonance branches, whose degeneracy is restored in the Lorentz-invariant
limit. Assuming that the matter vacuum-polarization tensor retains its
standard transverse form, we have shown that the two projector sectors remain
orthogonal under Dyson resummation. Their decay rates are determined
independently on the corresponding dispersion branches. We have also
emphasized that the displacement of the poles cannot be resolved by treating
the Lorentz-violating quadratic operator through only a finite number of
perturbative insertions; the exact propagator, or an equivalent all-order
resummation, is required.

For fermion annihilation, the contribution associated with the second
resonance is strongly controlled by the initial momentum geometry. In
particular, the explicit pole-splitting terms contain the common kinematic
factor $n^{2}k^{2}-4p_{1n}p_{2n}$. For a timelike preferred direction and
massless head-on incoming fermions, this factor vanishes identically, so that
the remaining Lorentz-violating dependence is contained exclusively in the
first resonance factor. By contrast, a highly boosted configuration with
nearly parallel incoming momenta can produce corrections enhanced as
$E^{2}/M^{2}$ and, for the quadratic pole-splitting contribution, as
$E^{4}/M^{4}$. Such a configuration should be regarded as a kinematic
benchmark rather than as the standard leading-order geometry of a hadron collider.

For spacelike and lightlike preferred directions, the exact timelike
cancellation is not generic. The second-branch contribution becomes direction
dependent and, in an Earth-based experiment, may exhibit sidereal modulation.
In the standard head-on configuration, however, the energy enhancement is
reduced and only the $\delta^{2}$ term receives $E^{2}/M^{2}$ amplification,
indicating that this geometry is not favorable for obtaining a large rate
enhancement from the second pole.

The resulting phenomenology should therefore not generally be understood as
two easily resolvable peaks. Since the splitting is controlled by the small
Lorentz-violating parameter, the two branches are expected to produce an
overlapping resonance structure whose observable manifestation depends on the
polarization content, momentum configuration, and orientation relative to the
preferred direction. Extending the analysis to realistic electroweak
couplings, non-Abelian vector fields, and collider production distributions
would be a natural next step.

\section*{Acknowledgments}

The author thanks Jon Chkareuli and Juansher Jejelava for useful discussions.

\end{document}